\documentclass[apj]{emulateapj}
\usepackage{amssymb,amsmath}
\usepackage{graphicx}
\usepackage{natbib}
\usepackage[OT1,T1]{fontenc}

\newcommand{\ha}{H$\alpha$}
\newcommand{\ca}{\ion{Ca}{2} 8542 \r{A}}
\newcommand{\kms}{km s$^{-1}$}
\newcommand{\dlb}{$\delta\lambda_{b}$}

\begin{document}

\title{Spectral observations of Ellerman bombs and fitting with a two-cloud model}
\author{Jie Hong\altaffilmark{1,2}, M.~D. Ding\altaffilmark{1,2}, Ying Li\altaffilmark{1,2}, Cheng Fang\altaffilmark{1,2}, and Wenda Cao\altaffilmark{3}}
\affil{\altaffilmark{1}School of Astronomy and Space Science, Nanjing University, Nanjing 210093, China \email{dmd@nju.edu.cn}}
\affil{\altaffilmark{2}Key Laboratory for Modern Astronomy and Astrophysics (Nanjing University), Ministry of Education, Nanjing 210093, China}
\affil{\altaffilmark{3}Big Bear Solar Observatory, New Jersey Institute of Technology, 40386 North Shore Lane, Big Bear City, CA 92314-9672, USA}

\begin{abstract}
  We study the H$\alpha$ and \ion{Ca}{2} 8542 \r{A} line spectra of four typical Ellerman bombs (EBs) in active region NOAA 11765 on 2013 June 6, observed with the Fast Imaging Solar Spectrograph installed at the 1.6 meter New Solar Telescope at Big Bear Solar Observatory. Considering that EBs may occur in a restricted region in the lower atmosphere, and that their spectral lines show particular features, we propose a two-cloud model to fit the observed line profiles. The lower cloud can account for the wing emission, and the upper cloud is mainly responsible for the absorption at line center. After choosing carefully the free parameters, we get satisfactory fitting results. As expected, the lower cloud shows an increase of the source function, corresponding to a temperature increase of 400--1000 K in EBs relative to the quiet Sun. This is consistent with previous results deduced from semi-empirical models and confirms that a local heating occurs in the lower atmosphere during the appearance of EBs. We also find that the optical depths can increase to some extent in both the lower and upper clouds, which may result from either a direct heating in the lower cloud, or illumination by an enhanced radiation on the upper cloud. The velocities derived from this method, however, are different from those obtained using the traditional bisector method, implying that one should be cautious when interpreting this parameter. The two-cloud model can thus be used as an efficient method to deduce the basic physical parameters of EBs.
\end{abstract}


\section{Introduction}
Ellerman bombs (EBs) are small-scale solar activities that occur in or near active regions. A striking feature of EBs is emission in \ha\ line wings that was first observed by \citet{1917ellerman}. EBs are also observed in other wavelength bands including \ca\ \citep{2006fang,2006socas,2007pariat,2013vissers,2013yang}, \ion{Ca}{2} H and K lines \citep{2008matsumotob,2010hashimoto,2013nelsonb}, G band \citep{2011herlender,2013nelsona}, and ultraviolet bands like AIA 1600 and 1700 \r{A} and TRACE 1600 \r{A} \citep{2007pariat,2013vissers,2013nelsona,2000qiu,2010berlicki}. The lifetime of a typical EB is considered to be about 5--10 min \citep{1973roy,1982kurokawa,1987zachariadis,1998nindos,2000qiu,2010hashimoto}.

A typical \ha\ line of EBs has three components, a central absorption, a Gaussian emission core, and a power-law wing \citep{1983kitai}. Usually, EBs are first seen in the blue wing, and then visible in both wings \citep{2010herlender}. There is also an asymmetry in EB line profiles. \citet{2006fang} found that for most EBs, the intensity of the blue wing is stronger than that of the red wing. The line asymmetry can change with time. \citet{2010hashimoto} found a relationship between the blue asymmetry of the \ion{Ca}{2} H line and the subcomponents of EBs seen in filtergrams. Such line asymmetries reflect mass motions at the EB formation heights. Usually, a bisector method was used to derive the mass motion velocity from the asymmetric contrast profiles. This method yields an upward velocity of less than 10 \kms\ in the lower chromosphere \citep{2008matsumotoa,2013yang}. In addition, a downward velocity of less than 1 \kms\ was also found in the photosphere \citep{2002georgoulis,2008matsumotoa,2013yang}.

The structure of EBs has been studied using high resolution filtergrams. EBs are elliptically shaped, with a bright core of about $0.7\arcsec\times0.5\arcsec$ and a diffuse halo of about $1.2\arcsec\times1.8\arcsec$ in size \citep{1982kurokawa,2007pariat,2008matsumotob}. An EB can be further divided into subcomponents with an aspect ratio of 2.7 \citep{2010hashimoto}. Besides, EBs are related to moat flows and expansion of granular cells \citep{1997dara,2011watanabe,2013vissers,2013yang}.

Non-LTE models have been used to infer the formation height and temperature increase of EB regions. It was proposed that EBs occur at a place low in the atmosphere, i.e., somewhere in the lower chromosphere, the temperature minimum region, and the upper photosphere \citep{2006fang,2006socas,2010berlicki,2013gonzalez}. To account for the observed \ha\ and \ca\ lines, a temperature increase of 600--1300 K at the EB occurrence site is required for thermal models, and a value somewhat lower for nonthermal ones \citep{2006fang}. Sometimes, a higher temperature increase, say, 3000 K, was deduced \citep{2010berlicki}.

A number of numerical simulations have suggested that EBs are caused by magnetic reconnection in the lower atmosphere. A possible scenario is that emerging magnetic loops expand and reconnect with each other, resulting in a local heating \citep{2007isobe,2009archontis}. Magnetic field measurements confirmed that most EBs lie near the polarity inversion line \citep{1997dara,2006fang,2007pariat,2008watanabe,2013vissers,2013yang,2013gonzalez}, and are associated with magnetic flux emergence. Note that a small number of EBs appear in the unipolar regions, implying reconnection in a shearing magnetic field \citep{2008watanabe,2013vissers}.

In this paper, we investigate the physical parameters of EBs, using a two-cloud model. The paper is organized as follows. In Section 2, we introduce the observations and the selected EBs for study. Section 3 presents a preliminary spectral analysis based on the bisector method. We then introduce our two-cloud model and show the fitting results in Section 4, followed by a discussion and conclusion in Section 5.

\section{Observations}
The observations were performed using the Fast Imaging Solar Spectrograph (FISS) of the 1.6 meter New Solar Telescope (NST, \citealt{2012goode,2010cao}) at Big Bear Solar Observatory. The NST, outfitted with a high-order adaptive optics system that has 308 sub-apertures over the telescope pupil, and state-of-the-art focal plane instruments, is the first facility-class solar telescope built in the U.S. in a generation. FISS is an imaging spectrograph that adopts an Echelle disperser with field-scanning method, with which two-dimensional spectra and images of dual spectral bands (\ha\ and \ca) can be acquired simultaneously \citep{2013chaea}. FISS can probe into the physical processes in the photosphere and chromosphere, including plasma flows and oscillations in various phenomena, like fibrils, prominence, and EBs \citep{2013andic,2013chaeb,2013cho,2013maurya,2013park,2013yang}. Using FISS, we obtained \ha\ and \ca\ spectra with high spatial and spectral resolution for a target area near the pores in NOAA 11765 on 2013 June 6. The spectral resolution is 19 m\r{A} for \ha, and 26 m\r{A} for \ca. The scanning lasted for nearly 1 hour, from 16:46:39 to 17:39:49 UT, with a time cadence of 27 s, covering a field of view (FOV) of 40\arcsec\ in the slit direction by 24\arcsec\ in the scan direction. The seeing was nearly 1\arcsec\ during the observations.

\begin{figure}
    \epsscale{0.7}
    \plotone{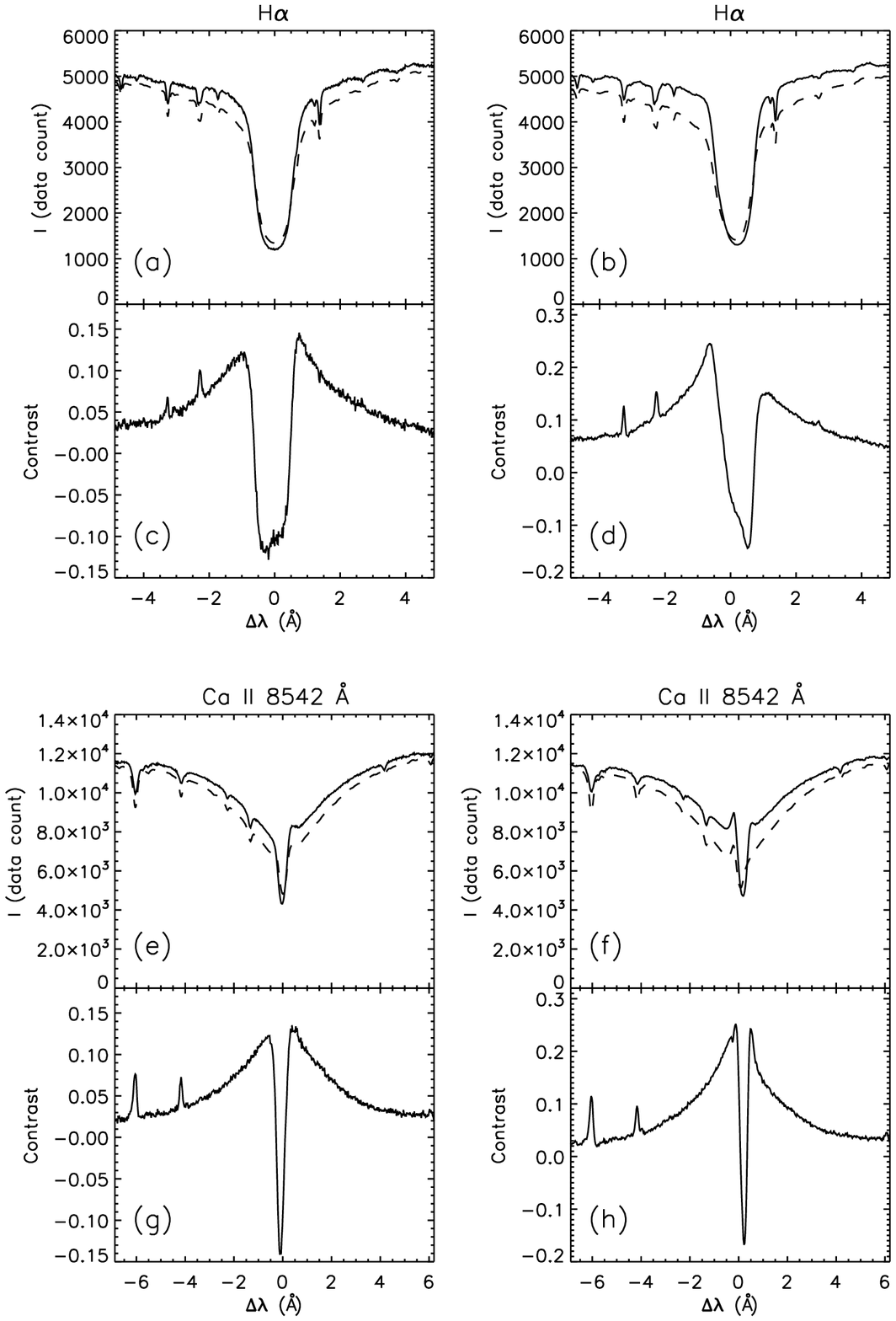}
    \caption{Line profiles of \ha\ and \ca\ for EB1 (left column) and EB3 (right column). (a)-(b) \ha\ line profiles of the EB (solid) and the nearby quiet Sun (dashed). (c)-(d) The contrast profiles of \ha. (e)-(h) Same as (a)-(d), but for the \ca\ line.}\label{vp}
\end{figure}

From the reconstructed images and line profiles in the observed region, we identified over 20 EBs, with the feature of excess emissions in line wings. Most of these EBs are relatively small and weak. Among them, we selected four typical EBs (hereafter called EB1, EB2, EB3, and EB4, respectively) with obvious emission features, each of which having a size of about 2\arcsec. Figure~\ref{vp} shows the \ha\ and \ca\ lines for two EBs, in which the contrast profiles are defined as
\begin{equation}
C(\lambda)=\frac{I_{EB}-I_{q}}{I_{q}},
\end{equation}
where $I_{EB}$ is the spatially averaged profile of each EB, and $I_{q}$ is the mean profile of the quiet Sun. The latter is averaged over an area near the corresponding EB at the same time to reduce the random noise. Note that, there exist some photospheric absorption lines as well as terrestrial lines in the \ha\ and \ca\ line wings. Here, these blended lines have not been removed from the observed spectra, which cause some pseudo emission spikes in the contrast profiles (Figure~\ref{vp}). From the figure, one can see clearly the three components of an EB contrast profile as noted by \citet{1983kitai}, namely, a central absorption, a Gaussian core, and a power-law wing. Another typical feature is the line profile asymmetry. As examples, EB1 shows a stronger red wing while EB3 shows a stronger blue wing, which could imply different dynamical processes or different emission/absorption features in them. Note that in the quiet region near EB3, the \ha\ line has a slight shift at the central part that may be caused by some flows in the chromosphere.

Among the four selected EBs, EB3 is the strongest; it lasted more than half an hour and did not fade out at the end of our observations. EB4 is the weakest and it moved out of the FOV at a later time. EB1 and EB2 are intermediate; EB1 showed the whole evolution, while EB2 was out of the FOV at the beginning but later on moved into it. Note that the change of the relative positions of the EBs in the FOV was totally due to the slight movement of the FOV itself. In addition, the typical feature of EB4 was not prominent during 17:03:59--17:09:55 UT due to the disturbance of the air, so that the data in this period are not used in our analysis.

\section{Preliminary spectral analysis}

\subsection{The \ha\ and \ca\ lines}
From the line profiles of \ha\ and \ca, we can derive mass flow velocities in the EBs based on the Doppler shifts. Before this, some terrestrial and absorption lines should be removed carefully.

\begin{figure}
    \epsscale{1}
    \plotone{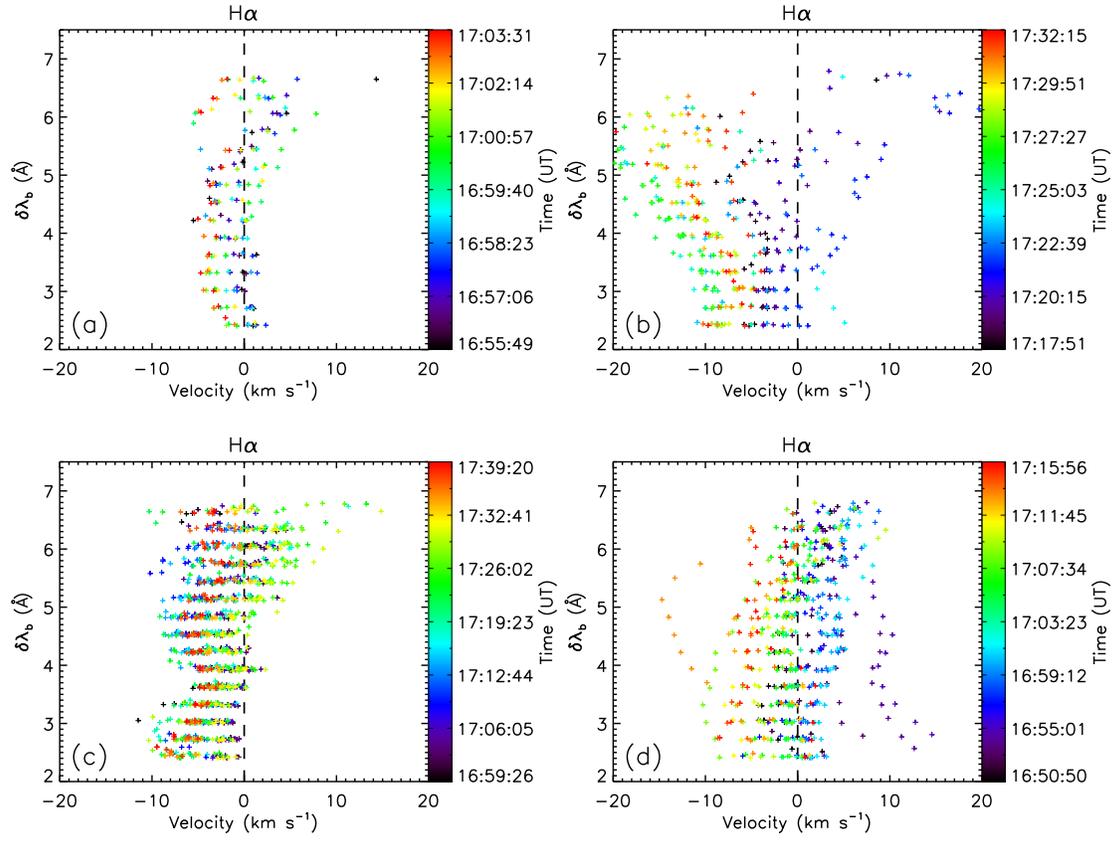}
    \caption{Velocities derived from the bisector with different values of \dlb\ for the \ha\ line. Positive velocities denote redshifts, implying downflows. Panels (a), (b), (c), and (d) are for EB1, EB2, EB3, and EB4, respectively.}\label{va1}
\end{figure}

\begin{figure}
    \plotone{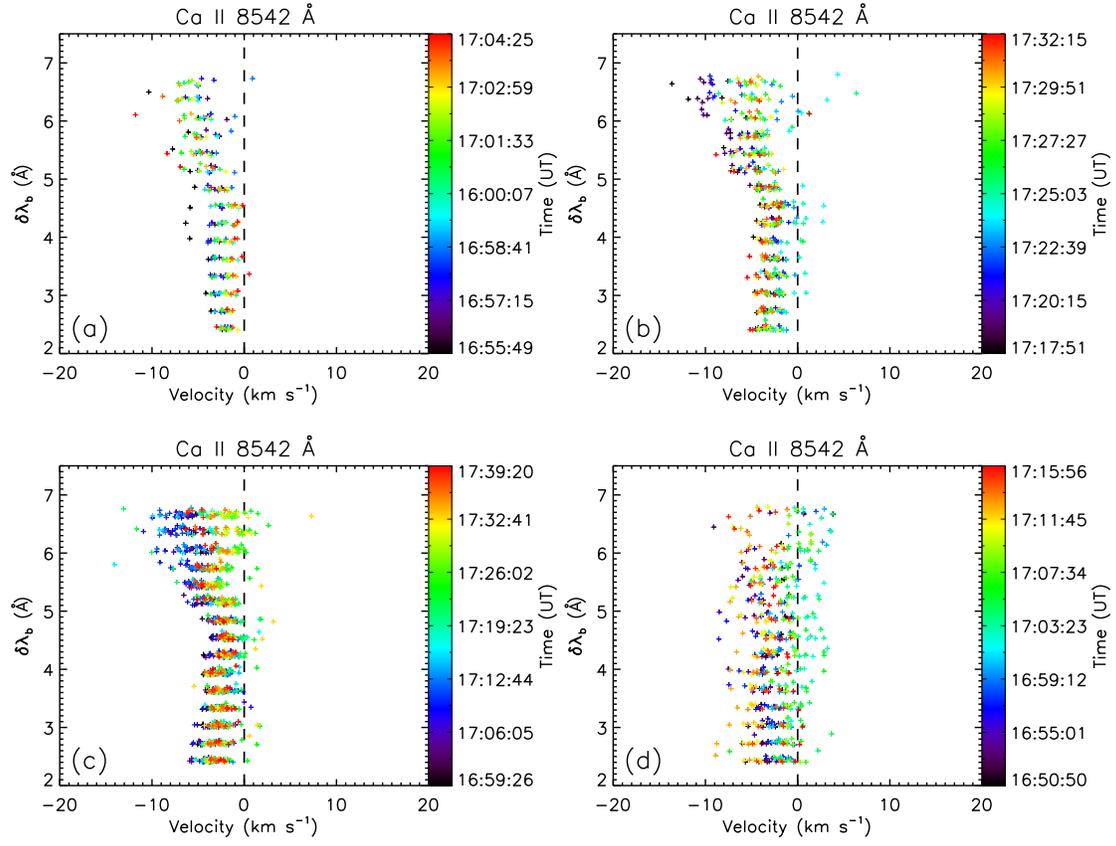}
    \caption{Same as Figure~\ref{va1}, but for the \ca\ line.}\label{vb1}
\end{figure}

A simple method to derive the velocities from the emission components of the contrast profiles is the bisector method, also referred to as lambdameter. This method makes a horizontal cut of a line with a certain full width \dlb\ so that
\begin{equation}
C\left(\lambda_{m}-\frac{\delta\lambda_{b}}{2}\right)=C\left(\lambda_{m}+\frac{\delta\lambda_{b}}{2}\right),
\end{equation}
where $\lambda_{m}$ is considered to be the observational line center of the emission component. Velocities are then derived from the Doppler shift of the line center relative to the theoretical (rest) one. In fact, different values of \dlb\ represent different emission levels, which may yield different Doppler shifts. For the \ha\ line, \citet{2013yang} chose \dlb\ to be 2.8 \r{A} and thought that it can represent the emission characters well. \citet{2008matsumotoa}, however, chose 20 different intensity levels and calculated the averaged line center. To show how the velocities vary with the emission levels, we plot in Figures~\ref{va1} and \ref{vb1} the velocities deduced from the \ha\ and \ca\ contrast profiles as a function of the parameter \dlb. The time evolution is also shown using different colors. It is known that different parts of a line profile are formed at different layers. For example, the \ha\ line center is formed in the upper chromosphere while the line wing originates from the photosphere. However, the dynamic process of an EB may be restricted to a small height range. This explains why the velocities deduced above can vary with emission levels not only in magnitude but also sometimes in sign.

To quantify this difference, we choose two different values of \dlb\ (3.1 \r{A} and 5.2 \r{A} for the \ha\ line in EB1 and EB4; and 3.1 \r{A} and 6.1 \r{A} for the \ha\ line in EB2 and EB3, as well as for the \ca\ line in all the four EBs) and plot the deduced velocities of the EBs as shown in the top two rows of Figure~\ref{veb}. For the \ha\ line, if we choose a smaller \dlb, we can get an upward velocity of 0--5 \kms\ from the \ha\ line, which is consistent with previous results \citep{2008matsumotoa,2013yang}. However, in most cases, a larger \dlb\ likely yields a downward velocity. This might be explained by the position of magnetic reconnection. A smaller \dlb\ may correspond to a layer relatively high, say, above the reconnection point; while a larger \dlb\ may correspond to a layer beneath the reconnection point. Then, different velocity signs are possible if explained in terms of different reconnection outflows. It is noticed that for the \ca\ line, the difference in velocities derived from different \dlb\ values are less obvious than \ha\ as shown in Figure~\ref{va1}. This might be due to the narrower line width of \ca\ compared with \ha. Note that the bisector method is only an approximate one since it does not consider the radiative transfer effect that is crucial for optically thick lines. Theoretically, the velocities derived at the far wings are more reliable since it originates from a relatively small height range than that from the line center.  However, the far wings suffer from the influence of noise and some blanketing lines that, on the contrary, debases the reliability of the results. Therefore, choosing a suitable emission level is a key factor when using the bisector method.

\subsection{The \ion{Ti}{2} line}
The absorption line \ion{Ti}{2} 6559.567 \r{A} is formed in the photosphere. This line is relatively simple and can be fitted directly by a Gaussian profile. In this way, we can obtain the velocity in the photosphere. Note that we have removed the effects of solar oscillations and solar rotation. The results are shown in the third row of Figure~\ref{veb}. A downward flow of 0--0.4 \kms\ is found in the photosphere of the EBs, which is consistent with previous results \citep{2002georgoulis,2008matsumotoa,2013yang}. This result shows that the EBs occur somewhere above the formation height of the \ion{Ti}{2} line if the downward flow is interpreted as the reconnection outflow.

\section{Spectral fitting with a two-cloud model}

\subsection{Two-cloud model}
\begin{figure}
    \plotone{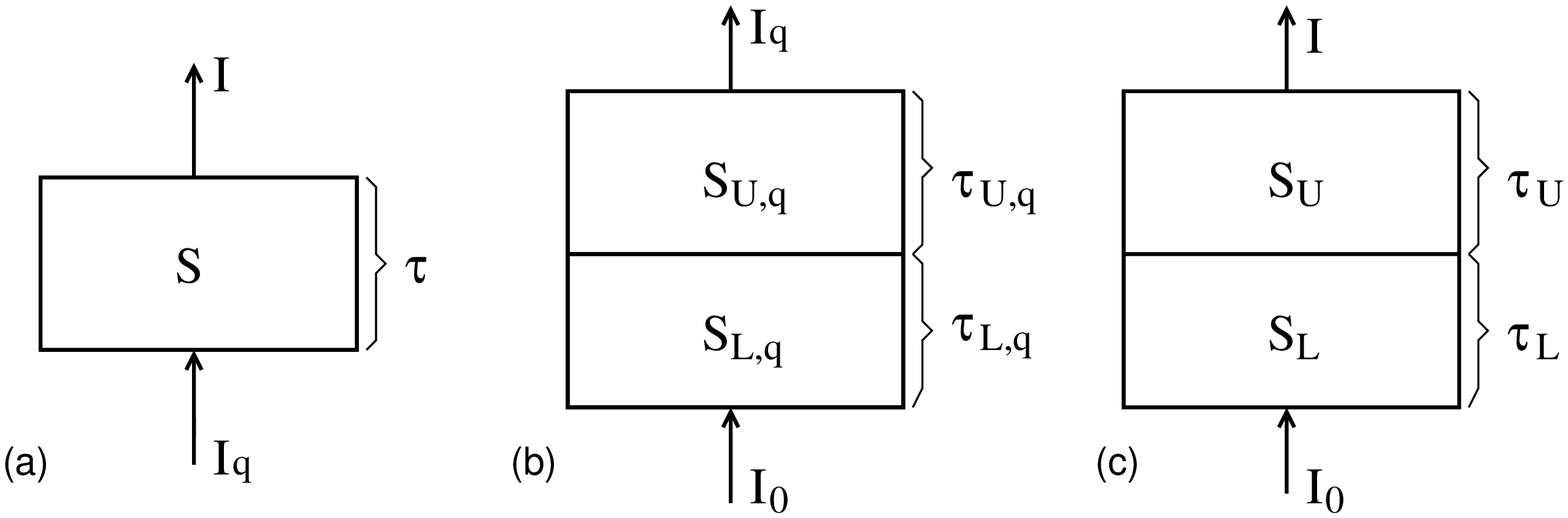}
    \caption{Schematic models for spectral fitting. (a) Beckers' cloud model. (b) The two-cloud model used for the quiet Sun. (c) The two-cloud model used for the EB.}\label{cm}
\end{figure}

The traditional cloud model was first proposed by \citet{1964beckers}. This model treats an active object as a cloud above the photosphere (Figure~\ref{cm}(a)). The parameters of the cloud are then inferred by fitting the line profiles emergent from the top of the cloud with the observed ones. Denoted by $I_{q}$, the intensity from the quiet Sun, and by $I$ that from the active region under study, can be related by solving the radiative transfer inside the cloud,
\begin{equation}
I=I_{q}\exp(-\tau)+S[1-\exp(-\tau)],
\end{equation}
where $S$ and $\tau$ are the source function and optical depth of the cloud, respectively. If the line profile is only subject to Doppler broadening, then the optical depth can be expressed as a Gaussian profile.
The contrast profile is then derived as
\begin{equation}
C=\left(\frac{S}{I_{q}}-1\right)[1-\exp(-\tau)].\label{bcm}
\end{equation}

In Beckers' cloud model, the cloud-like structure of plasma is located above the background, and the incident light at the bottom of the cloud is considered the same as the observed intensity at the surface of the background. Recently, \citet{2014chae} proposed an embedded cloud model that is more representative for real cases. In this model, the cloud is embedded in the background, so that the incident light on the cloud comes from some intrinsic layer of the background, which is not the same as that emergent from the surface.

There are only four free parameters in Beckers' cloud model, namely, the source function, optical depth at line center, wavelength of line center, and Doppler width. From Equation~(\ref{bcm}), it is clearly seen that no matter how one varies the value of $S$, the contrast profile is either in emission or in absorption. In an EB, however, a certain layer in the lower atmosphere is heated while the upper chromosphere is almost undisturbed. Therefore, the particular layer of EB occurrence can be emissive relative to the quiet Sun but the upper layers can still keep absorptive. This feature is reflected in the contrast profile that exhibits an emission wing and an absorption core. Owing to this diversity, the traditional cloud model within which all the parameters are taken as constant cannot fully reproduce a typical EB line profile. Multiple cloud components with different emission/absorption features are then required. Here, we propose a two-cloud model, in which the lower cloud refers to the possible heating region, and the upper cloud represents the less disturbed upper atmosphere.

The methodology of the two-cloud model is described as follows. As illustrated in Figure~\ref{cm}(b)-(c), assuming that the intensity illuminating the lower cloud from below is $I_{0}$, the observed intensity $I$, emergent from the top of the upper cloud, can then be expressed as
\begin{equation}
\begin{aligned}
I&=I_{0}\exp[-(\tau_{L}+\tau_{U})]+S_{L}[1-\exp(-\tau_{L})]\exp(-\tau_{U})\\
&+S_{U}[1-\exp(-\tau_{U})],\label{tcm}
\end{aligned}
\end{equation}
where the subscripts ``L'' and ``U'' denote parameters in the lower and upper clouds, respectively. Generally speaking, the main broadening mechanisms for the \ha\ line are radiative damping and Doppler broadening; therefore, the absorption coefficient should have a Voigt profile. A Voigt profile is a convolution of a Gaussian profile, which contributes mainly to the line core, and a Lorentzian profile, which contributes mainly to the line wings. Since the upper cloud is mainly responsible for the absorption core, for simplicity, we assume that it is only subject to Doppler broadening; therefore, the optical depth of the upper cloud has the following form:
\begin{equation}
\tau_{U}=\tau_{U}^{0}\exp\left[-\left(\frac{\lambda-\lambda_{U}}{\Delta\lambda_{D}}\right)^{2}\right],
\end{equation}
where $\tau_{U}^{0}$ is the optical depth at line center, $\lambda_{U}$ is the observed wavelength of line center, and $\Delta\lambda_{D}$ is the Doppler width that can be expressed as
\begin{equation}
\Delta\lambda_{D}=\frac{\lambda_{0}}{c}\sqrt{\frac{2kT}{m}+\xi^{2}}.
\end{equation}
In the equation above, $\lambda_{0}$ is the rest wavelength of line center, $c$ is the speed of light, $T$ is the kinetic temperature, $m$ is the atomic mass, and $\xi$ is the velocity of microturbulence. Similarly, as the lower cloud contributes mainly to the emission wing, we assume that it is only subject to radiative damping as described with a Lorentzian profile:
\begin{equation}
\tau_{L}=\tau_{L}^{0}\frac{\delta^{2}}{\delta^{2}+(\lambda-\lambda_{L})^{2}},
\end{equation}
where $\delta$ is the damping constant, and $\tau_{L}^{0}$ and $\lambda_{L}$ have similar meanings.

It should be noted that, in our model, the quantity $I_{q}$, the quiet-Sun intensity emergent from the top of the upper cloud, is different from $I_{0}$, the intensity incident on the bottom of the lower cloud. Such a treatment is different from the traditional cloud model but similar to the embedded cloud model proposed by \citet{2014chae}. Although the two clouds in our model are used to account for the EB spectra, they do exist in quiet regions with, however, different parameters, as depicted in Figure~\ref{cm}(b). For a discrimination, the second subscript ``q'' refers to parameters in the quiet region.

\subsection{Pre-processing}
The contrast profile, derived with the two-cloud model, is a rather complicated expression. We have to reduce the numbers of free parameters so that the fitting can become practical. This is performed through several steps. First, we fix some parameters considering their specific physical meanings.

When we apply Equation (\ref{tcm}) to the quiet region, we can assume that the two clouds are static, namely, $\lambda_{L,q}=\lambda_{U,q}=\lambda_{0}$, where $\lambda_{0}$ is the rest wavelength of line center. Besides, what is interesting is not the absolute values of the parameters, but the relative changes of them from the quiet region to the EBs. We thus introduce three parameters, $\alpha_{L}$, $\alpha_{U}$, and $\beta$, to quantify such changes as $\tau_{L}^{0}=(1+\alpha_{L})\tau_{L,q}^{0}$, $\tau_{U}^{0}=(1+\alpha_{U})\tau_{U,q}^{0}$, and $S_{L}=(1+\beta)S_{L,q}$. Here, we neglect the change of the source function of the upper cloud considering there is little heating in the upper layers. Similarly, the Doppler width of the upper cloud, related to the temperature and microturbulence there, is assumed not to change. For the lower cloud, the radiative damping constant is mainly dependent on the radiation field from below; thus, it can also be taken as a constant. In summary, there remain eleven unknown parameters in our two-cloud model, namely, $S_{L,q}$, $\tau_{L,q}^{0}$, $\lambda_{L}$, $\delta_{q}$, $S_{U,q}$, $\tau_{U,q}^{0}$, $\lambda_{U}$, $\Delta\lambda_{D,q}$, $\alpha_{L}$, $\alpha_{U}$, and $\beta$.

The next attempt is to find out typical values for those parameters in the quiet region (with the subscript ``q'') that are assumed not to change. We describe those parameters below.

\paragraph{Source function}
Under the assumption of complete frequency redistribution, the line source function for \ha\ can be calculated as
\begin{equation}
S_{\lambda}=\frac{2hc^{2}}{\lambda^{5}}\frac{1}{\frac{b_{2}}{b_{3}}\exp\left(\frac{hc}{\lambda kT}\right)-1}.\label{func}
\end{equation}
In Equation(\ref{func}), the departure coefficients of the hydrogen atom at energy levels 2 and 3, $b_{2}$ and $b_{3}$, and the temperature, $T$, can be taken from the VAL-C quiet-Sun model \citep{1981vernazza}. Since the EBs are supposed to occur from the upper photosphere to the lower chromosphere, we choose the corresponding layers in the VAL-C model that are at the height range of 300--700 km to calculate the parameter range of $S_{L,q}$. The parameter range of $S_{U,q}$ is calculated at the height range of 1500--2000 km to represent chromospheric features.

\paragraph{Doppler width and damping constant}
These two parameters are related to the width of the corresponding profile. In practice, we use the shape of the line wings to estimate the parameter range of $\delta_{q}$, and the shape of the absorption core for the parameter range of $\Delta\lambda_{D,q}$. The primary range is 1--5 for $\delta_{q}$, and 0.3--0.6 for $\Delta\lambda_{D,q}$, respectively.

\paragraph{Optical depth at line center}
This parameter is less known. Assuming that the EBs are restricted to a small height range, the lower cloud is likely optically thin. The upper cloud should be optically thick as reflected from the strong absorption in line core. We set primarily the range of $\tau_{L,q}^{0}$ to be 0.2--0.7 and the range of $\tau_{U,q}^{0}$ to be 1--3.

After setting each of the above six parameters a variation range as described above, we then search for the optimal line profile in the parameter space that can best match the observed one. This process results in an optimal set of parameters for the quiet region as well as the initial guess of other free parameters for the EBs. Here, we notice that $\tau_{L,q}^{0}$ is tightly coupled with $\alpha_{L}$, since only the product of them represents the opacity increase of the lower cloud that is crucial to fitting the EB profile. This means that an overestimate/underestimate of the former can be compensated by a smaller/larger value of the latter. The same is for $\tau_{U,q}^{0}$ and $\alpha_{U}$. We further find that a misestimate of $\tau_{L,q}^{0}$ and $\tau_{U,q}^{0}$ will not significantly influence the fitted results of the other three free parameters. Therefore, in practical fitting, we choose the values of $\tau_{L,q}^{0}$ to be 0.4 and $\tau_{U,q}^{0}$ to be 2, which lie in the middle of the parameter space mentioned above. After this, only five free parameters are left: $\lambda_{L}$, $\lambda_{U}$, $\alpha_{L}$, $\alpha_{U}$, and $\beta$. This makes the two-cloud model fitting feasible. The complete set of parameters of the two-cloud model is listed in Table~\ref{tcmp}.

\subsection{Fitting results}
For a multi-parameter fitting, it is very important to choose reasonable initial values otherwise the fitting process might not converge. The reasonable initial guess of the five free parameters have been obtained in the pre-processing procedure. Then, the procedure \verb"curvefit.pro" of Interactive Data Language, which uses a gradient-expansion algorithm to compute a non-linear least squares fit, is adopted in our spectral fitting.

To check the reliability of the model and the fitting results, we use the Monte-Carlo method to estimate the intrinsic errors of the fitted parameters. To do so, we first assign an artificial noise to each single point of the contrast profile, and we then reperform the fitting. This procedure is repeated 100 times for one contrast profile, and the standard deviation of the fitted parameters is considered to be the intrinsic errors. The standard deviation of the artificial noise is set to be three times the intensity fluctuations at the far wings.

\begin{figure}
    \plotone{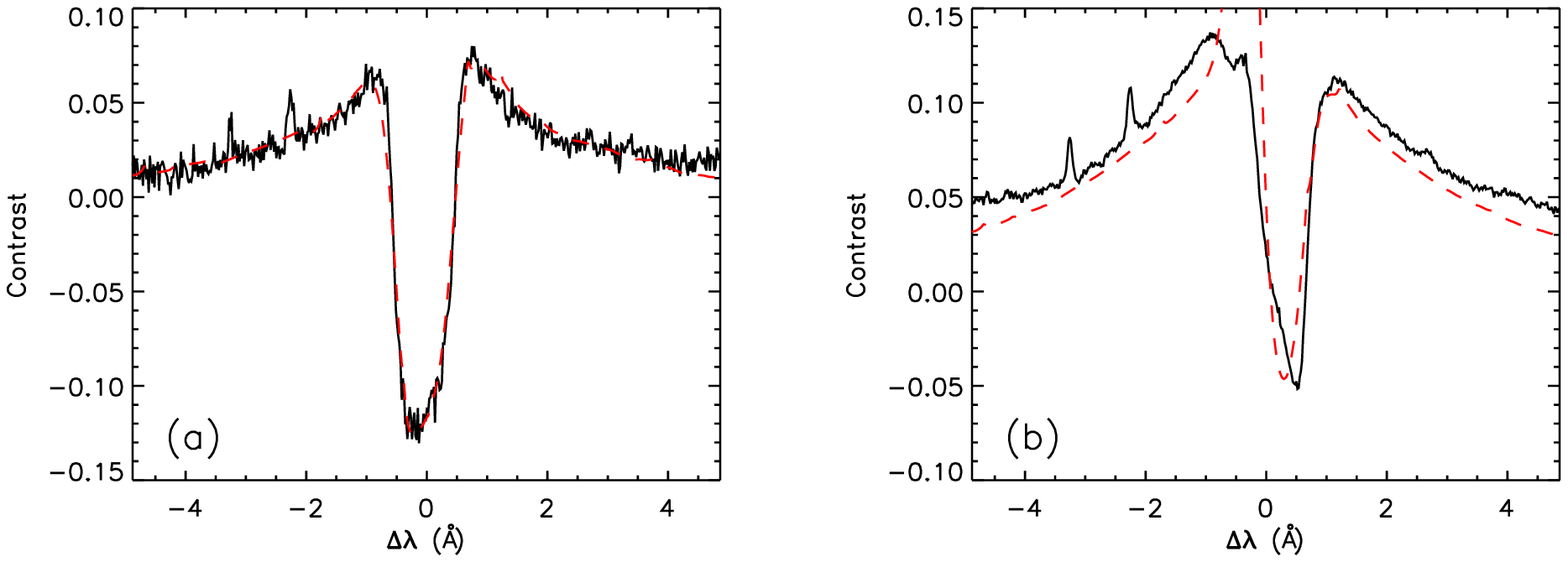}
    \caption{Typical contrast profiles of \ha\ (black) and fitting results (red) using the two-cloud model. The fitting is good in panel (a) but marginally acceptable in panel (b). About 80\% of the profiles have better fitting results than that in panel (b).}\label{fit}
\end{figure}

\begin{figure}
    \plotone{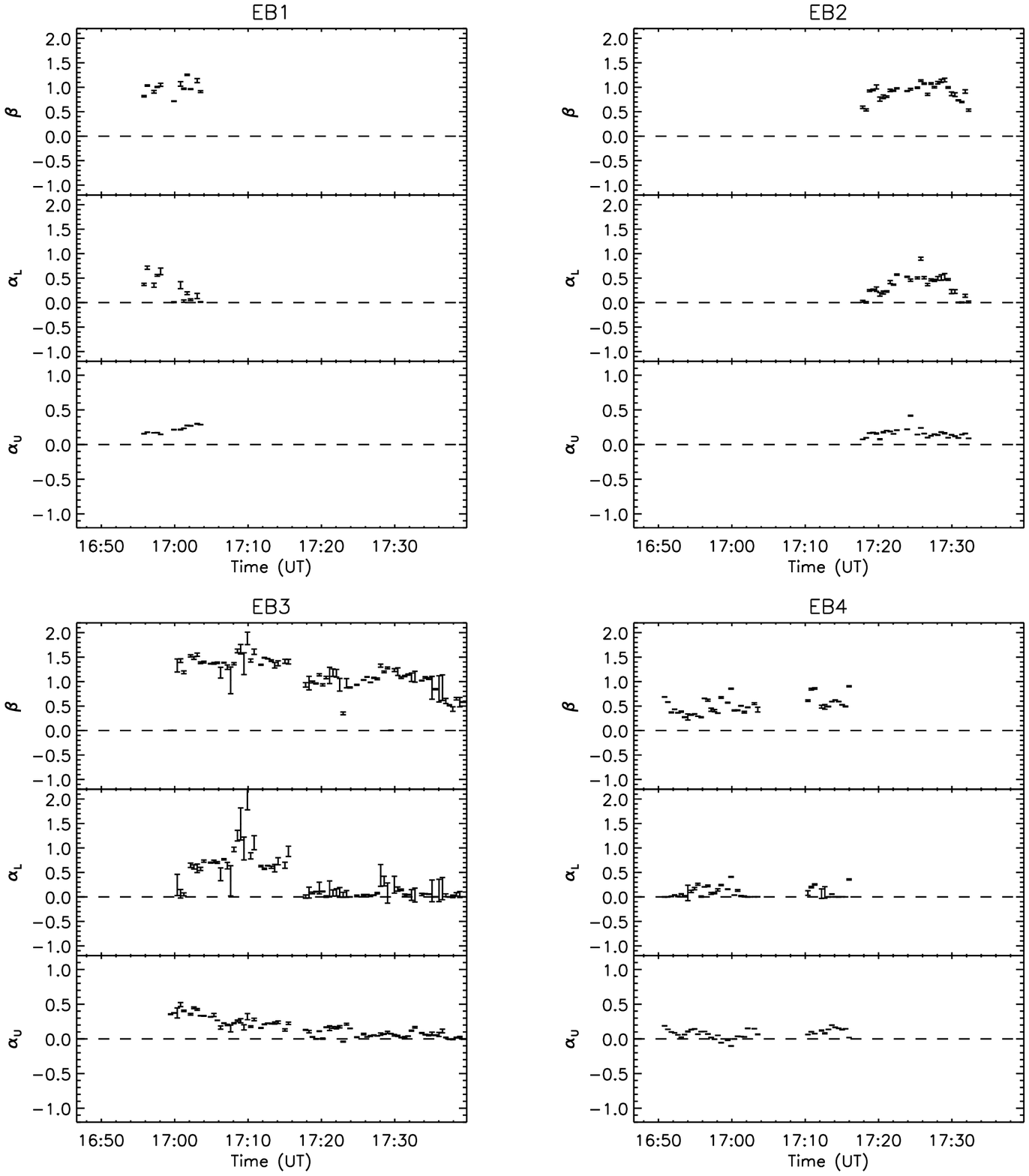}
    \caption{Parameters from the spectral fitting with the two-cloud model, showing an increase of source function in the lower cloud and an increase of opacity in both the lower and upper clouds in EBs. The error bars are computed from Monte-Carlo simulations.}\label{vpa}
\end{figure}

We plot some typical fitting results in Figure~\ref{fit} and show the values of $\alpha_{L}$, $\alpha_{U}$, and $\beta$ in Figure~\ref{vpa}. The velocities of the lower cloud and the upper cloud, $v_{L}$ and $v_{U}$, derived from the parameters $\lambda_{L}$ and $\lambda_{U}$, are shown in the bottom two rows of Figure~\ref{veb}.

A striking feature, yet expected, of the fitting results is an enhancement of the source function in the EBs with, however, different magnitudes. As shown in Figure~\ref{vpa}, the parameter $\beta$ can be as large as 2 for EB3, but is only about 0.5 for EB4. By applying Equation~(\ref{func}) to both the quiet Sun area and the EB regions, we can derive the temperature increase in the EB relative to the quiet Sun based on the increase of the source function. The results are shown in Table~\ref{tem}. It is seen that the temperature increase can be as large as about 1000 K for EB3, and only about 400 K for EB4. Note that this temperature increase is just for the EB occurrence layer (the lower cloud).

Besides, one can also find an obvious increase of the optical depth in both clouds. As shown in Figure~\ref{vpa}, for all the four EBs, the optical depths of the lower cloud increase by 10\%--50\%, while those of the upper cloud increase by 10\%--30\%. This implies an increase of the number density of the hydrogen excited levels mainly caused by a local heating in a lower layer but an enhanced radiation from below in upper layers when EBs occur.

It should be noted that the velocities from the two-cloud model fitting are quite different from the velocities derived from the bisector method. For the upper cloud, the velocities are less than 2 \kms, and tend not to vary dramatically. However, EB3 shows a very large downward velocity in the upper cloud. Probably, this is not EB-related, but caused by some other dynamical processes in the chromosphere. By comparison, the \ha\ line in the quiet region near EB3 also shows a red shift at the line center (Figure~\ref{vp}(b)).

\section{Discussion and conclusion}
In this paper, we for the first time use a two-cloud model to fit the EB profiles. The two-cloud model consists of two vertically adjacent clouds that possess different parameters, or emission/absorption features, in the solar atmosphere. Therefore, the most significant advantage of our two-cloud model is its capability to fit well both the absorption feature at the line center and the emission feature in the line wings, whereas the one-cloud model can only account for either of them. On the other hand, semi-empirical models are more sophisticated, and can yield height-varying parameters \citep{2006fang,2006socas,2014berlicki}; however, the inversion sometimes suffers from a relatively large uncertainty and requires more computational resources. By comparison, the two-cloud model yields only some selected (of course, most varying) parameters, yet it is sufficient in revealing the variation of the heated layer of EBs. Thus, the two-cloud model is more convenient and practical to deduce the basic physical parameters when dealing with a large sample of EBs.

Using the fitting results from the two-cloud model, we are able to estimate the temperature increase in the lower cloud where an EB appears. The temperature increase can vary from $\sim400$ to $\sim1000$ K in the four EBs. This is similar to the results of semi-empirical models showing a local temperature increase of 600--1300 K \citep{2006fang}. Our results confirm the local heating in the lower atmosphere of EBs. The heating in the lower cloud can also cause an increase of the optical depth of this cloud directly and that of the upper cloud indirectly through an enhanced illumination. Therefore, for a typical EB line profile, the emission in the line wings is primarily from the local heating in the lower cloud whereas the absorption core is due to the upper cloud whose optical depth is increased.

It is usually a difficult task to infer the Doppler velocities from the shifts and asymmetries of optically thick lines. The bisector method has been most often used. However, it is difficult to choose the width of the bisector (\dlb), or the emission level, as the formation of the line may cover a quite large height range and the occurrence height of EBs may differ from case to case. This can explain why the velocities derived using the bisector method vary so much. We confirm this point here. By comparison, spectral fitting with the two-cloud model can yield two velocities in the two clouds, which can be regarded as the average mass motions in these two layers. Thus, one should be cautious when comparing the results from the bisector method and the two-cloud method.

However, we should note some limitations of the two-cloud model. First, we have assumed that the absorption profile is either Gaussian (upper cloud) or Lorentzian (lower cloud). A more accurate treatment is to adopt a Voigt profile for both the clouds. Second, in the fitting, we have to fix some relatively less perturbed parameters. If more free parameters are used, we expect to reach better fitting to the line profiles, with, however,  larger uncertainties in fitted parameters owing to mathematical difficulties. The final point to note is that we use the mean profile of the EB region, i.e., we do not study the fine structures, if present, within each EB. Therefore, the deduced parameters only refer to average ones. It is possible that one gets a temperature increase somewhat higher than what we have deduced if using the spectra restricted to some particular points in EBs.

\acknowledgments
We are grateful to the BBSO team for their technical support for the observations and to the referee for valuable comments that helped improve the paper. This work was supported by NSFC under grant 11373023, and NKBRSF under grants 2011CB811402 and 2014CB744203. W. C. acknowledges the support of the US NSF (AGS-0847126) and NASA (NNX13AG14G).

\clearpage

\begin{figure}
    \plotone{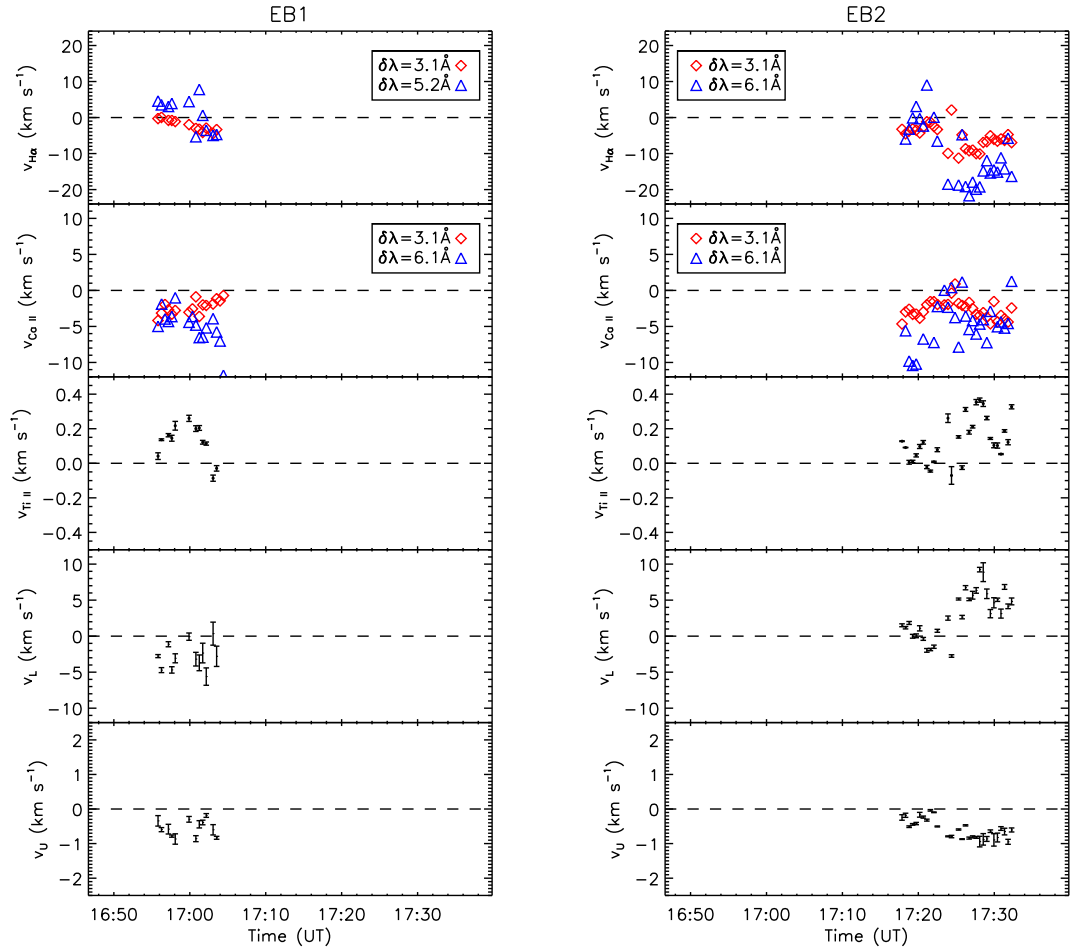}
    \caption{Various velocities derived for the four selected EBs. The top two rows show velocities derived from the bisector method for the two lines. Diamonds and triangles denote different values of \dlb. The third row is the photospheric velocity derived from the Doppler shift of the \ion{Ti}{2} line. The fourth row is for the velocity of the lower cloud, and the bottom one for the velocity of the upper cloud, derived from the two-cloud model fitting. The error bars in the bottom two rows are computed from Monte-Carlo simulations. The velocity sign has the same meaning as in Figure~\ref{va1}.}\label{veb}
\end{figure}

\begin{figure}
    \figurenum{\ref{veb}}
    \plotone{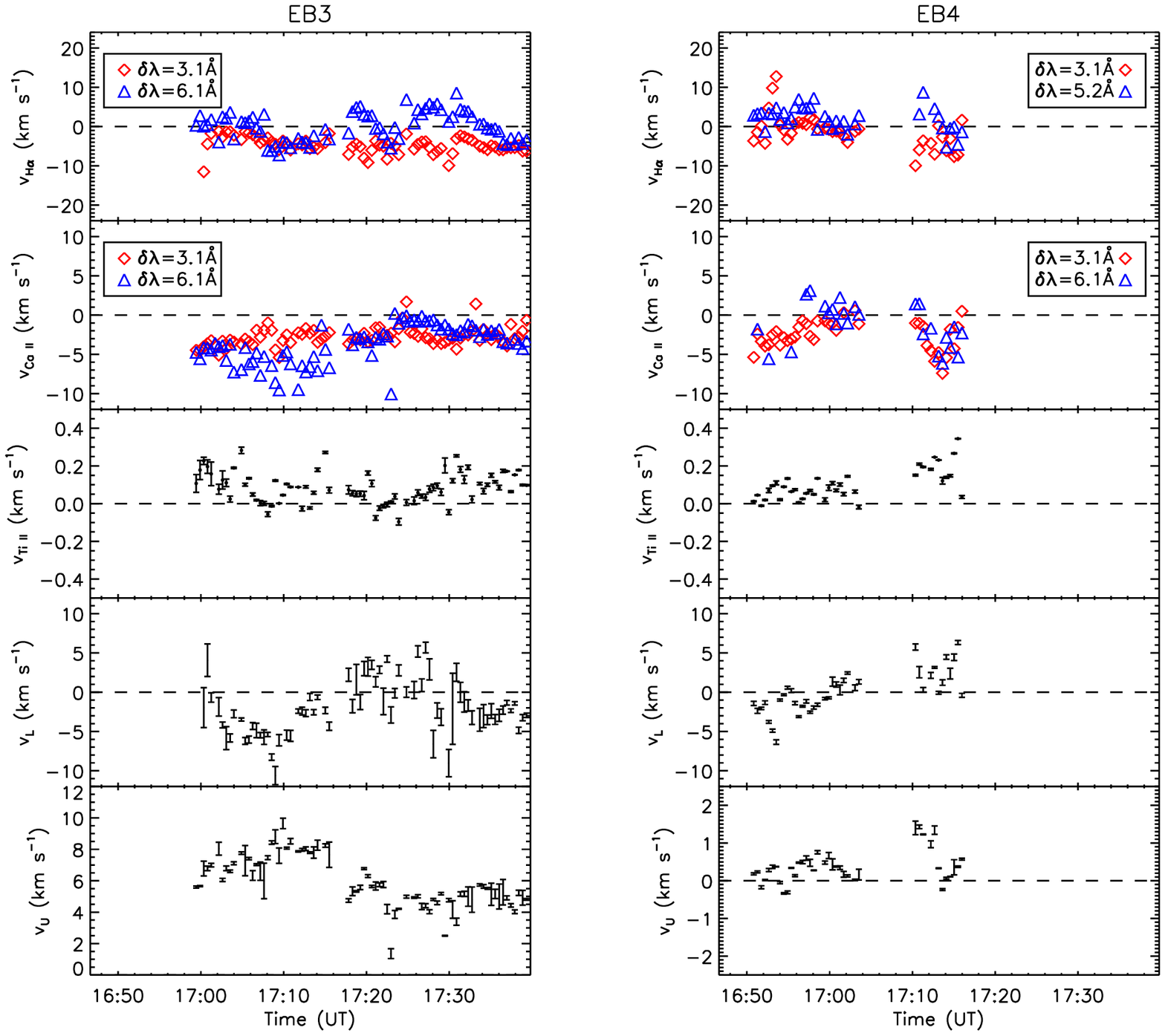}
    \caption{Continued.}
\end{figure}

\clearpage

\begin{deluxetable}{ccccc}
\tablecolumns{5}
\tablewidth{0pt}
\tablecaption{Parameters of the two-cloud model}
\tablehead{\colhead{Symbol} & \colhead{Definition} & \colhead{Cloud} & \colhead{Region} & \colhead{Treatment}}
\startdata
$S_{L,q}$ & Source function & Lower & Quiet Sun & Fixed\\
$S_{U,q}$ & Source function & Upper & Quiet Sun \& EB & Fixed\\
$\tau_{L,q}^{0}$ & Opacity at line center & Lower & Quiet Sun & Fixed\\
$\tau_{U,q}^{0}$ & Opacity at line center & Upper & Quiet Sun & Fixed\\
$\lambda_{L}$ & Wavelength at line center & Lower & EB & Free\\
$\lambda_{U}$ & Wavelength at line center & Upper & EB & Free\\
$\delta_{q}$ & Damping constant & Lower & Quiet Sun \& EB & Fixed\\
$\Delta\lambda_{D,q}$ & Doppler width & Upper & Quiet Sun \& EB & Fixed\\
$\alpha_{L}$ & Relative increase of $\tau_{L}^{0}$ & Lower & EB & Free\\
$\alpha_{U}$ & Relative increase of $\tau_{U}^{0}$ & Upper & EB & Free\\
$\beta$ & Relative increase of $S_{L}$ & Lower & EB & Free
\enddata\label{tcmp}
\end{deluxetable}

\begin{deluxetable}{ccc}
\tablecolumns{3}
\tablewidth{0pt}
\tablecaption{Increases of source function and temperature of the lower cloud in EBs}
\tablehead{\colhead{No.} & \colhead{$\beta$} & \colhead{$\Delta T$ (K)}}
\startdata
EB1 & 0.7-1.4 & 450-900\\
EB2 & 0.5-1.2 & 350-800\\
EB3 & 0.4-1.9 & 300-1200\\
EB4 & 0.2-0.9 & 150-700
\enddata\label{tem}
\end{deluxetable}

\end{document}